\documentclass[aps,prl,twocolumn,showpacs,superscriptaddress]{revtex4}
\usepackage{amsmath}
\usepackage{epsfig}
\usepackage{graphics,color}

\begin{document}
\title{Spin polarons and molecules in strongly-interacting atomic Fermi gases}
\author{P.~Massignan}
\affiliation{\mbox{Institute for Theoretical Physics, Utrecht University, Leuvenlaan 4, 3584 CE Utrecht, The Netherlands.}}
\author{G.~M.~Bruun}
\affiliation{Dipartimento di Fisica, Universit\`a di Trento and CNR-INFM BEC Center, I-38050 Povo, Trento, Italy.}
\affiliation{Niels Bohr Institute, University of Copenhagen, DK-2100 Copenhagen \O, Denmark.}
\author{H.~T.~C. Stoof}
\affiliation{\mbox{Institute for Theoretical Physics, Utrecht University, Leuvenlaan 4, 3584 CE Utrecht, The Netherlands.}}
\begin{abstract}
We examine pairing and molecule formation in strongly-interacting Fermi gases, and we discuss how radio-frequency (RF) spectroscopy can reveal these features.
 For the balanced case, the emergence of stable molecules in the BEC regime results in a two-peak structure in the RF spectrum with clearly visible medium effects on the low-energy part of the molecular wavefunction.
 For the highly-imbalanced case, we show the existence of a well-defined quasiparticle (a spin polaron) on both sides of the Feshbach resonance, we evaluate its lifetime, and we illustrate how its energy may be measured by RF spectroscopy.
\end{abstract}
\maketitle

 The experimental realization of strongly-interacting Fermi gases with Feshbach resonances constitutes one of the major recent breakthroughs in ultracold atomic physics. As a result, we can now study the properties of Fermi gases as a function of interaction strength, temperature, and population imbalance (polarization).
 A fundamental question is how we should understand pairing in such strongly-interacting systems.
 In the limit where attractions are weak and attractive, the BCS instability can lead to the formation of Cooper pairs, which are at the origin of our understanding of superconductivity.
 In the opposite Bose-Einstein condensation (BEC) limit, pairing leads to tightly bound molecules mixed with free atoms, which may be present due to either a nonzero temperature or polarization.
 In the intermediate regime close to the unitarity limit, the many-body physics is much more complicated.
 Radio-frequency (RF) spectroscopy represents at present the main experimental probe of strong correlations in ultracold atomic Fermi gases \cite{Regal,Chin}.
Recent experiments performed on a gas in the superfluid phase seem to be well described by a two-body theory with the molecule energy as a fitting parameter~\cite{Schunck08}.
 However, microscopic calculations for the unpolarized state indicate that both the superfluid and normal states at unitarity cannot be described simply as molecules mixed with free atoms~\cite{Perali,Massignan}.
 Furthermore, the highly polarized state remains normal even at $T=0$ and presumably is a Fermi liquid~\cite{Lobo}.

In this paper we investigate the nature of the strong
correlations, pairing, and molecule formation for polarized and
unpolarized atomic Fermi gases on both sides of the Feshbach
resonance.
 Our analysis concerns the normal phase, which arguably is not as well understood as the $T=0$ superfluid phase~\cite{Giorgini}, and we analyze how several nontrivial effects can be revealed in RF spectroscopy. 
 Pair fluctuations, which in the BCS and unitarity regimes lead to a pseudogap, change character in the BEC regime with the emergence of stable molecular states. We show how this change may be directly detected in the RF spectrum. For the highly-polarized case, we demonstrate the existence of a well-defined quasiparticle (spin-polaron) in the minority spectral function on both sides of the Feshbach resonance. We calculate its lifetime, and show how its energy may be measured via RF spectroscopy. We further discuss how
many-body effects significantly affect the molecular wavefunction and how this can be detected in RF spectroscopy. For a polarized gas, Pauli-blocking effects are shown to lead to an asymmetry between the minority and majority RF spectra. By comparing our results to a simple model of atoms mixed with molecules, we present clear physical interpretations 
of our results.
 Our analysis is based on a microscopic theory with no fitting parameters and, as we show, for high imbalance it compares excellently with recent Monte-Carlo results.

Consider a homogeneous Fermi mixture of two hyperfine states with
densities $n_\uparrow \ge n_\downarrow$. The interaction between
the two states is described by the $s$-wave scattering length $a$,
while the interactions between identical fermions may be ignored
due to the very low temperature of the gas.
 The chemical potentials $\mu_\sigma$ for the two components ($\sigma=\uparrow,\downarrow$) are determined from
$n_\sigma=-\partial\Omega/\partial{\mu_\sigma}$. We obtain the thermodynamic potential within the
Nozi\`eres and Schmitt-Rink (NSR) approximation~\cite{NSR} by writing $\Omega=\Omega_0+\Delta\Omega$, where $\Omega_0$ is the
ideal gas contribution and
\begin{equation}
\Delta\Omega =\!\int \! \frac {d \bf K}{(2\pi)^3} \! \int_{-\infty}^\infty \! \frac
{d\epsilon}{2\pi i} \frac{ \ln(1-T^{2B}\Xi^+)-\ln(1-T^{2B}\Xi^-)
}{\mathrm{e}^{\beta \epsilon}-1}.
\end{equation}
The two-body T-matrix is $T^{2B}=4\pi \hbar^2 a/m$,
 $\Xi^\pm=\Xi(K,\epsilon \pm i0^+)$ is the regularized propagator for the center-of-mass of two atoms with total momentum $\mathbf{K}$ and energy $\epsilon$, and $\beta=1/k_B T$.
 The single-particle properties are described by the spectral function
$A_\sigma(k,\omega)=-\textrm{Im}\{2/[\omega+i0^+-\xi_{k,\sigma}-\Sigma_\sigma(k,\omega+i0^+)]\}$,
where ${\mathbf{k}}$ and $\omega$ are the momentum and the energy of an atom in state $|\sigma\rangle$, and $\xi_{k,\sigma}=k^2/2m-\mu_\sigma$ is its kinetic energy measured with respect to its
chemical potential. To include the effects of interactions, we
calculate the selfenergy $\Sigma_\sigma(k,\omega+i0^+)$ within the ladder approximation.
 The above approach has been shown to compare well with Monte-Carlo calculations for thermodynamic and single-particle quantities~\cite{Pieri,Huliu,Lobo,Massignan}, especially in the highly-imbalanced case.

The properties of the atomic gases can be probed by RF photons with frequency $\omega_\mathrm{RF}$, which induce transitions from one of the hyperfine states,
 say $|\sigma\rangle$, to a third state $|e\rangle$ that is initially empty.
 Within linear response, the rate $\Gamma_{\sigma\rightarrow e}(\omega_\mathrm{RF})$
 is given by the imaginary part of the
Fourier transform of the retarded spin-flip correlation function
$-i\theta(t-t')\langle[\psi_e^\dagger({\mathbf{r}},t)
\psi_\sigma({\mathbf{r}},t),\psi_\sigma^\dagger
({\mathbf{r}}',t')\psi_e({\mathbf{r}}',t')]\rangle$. A conserving
calculation of this correlation function is a complicated problem \cite{Perali,Mueller}.
However, a major simplification arises for the spin mixture used in recent experiments, where final-state interactions between atoms in states
$|e\rangle$ and $|-\sigma\rangle$ can be ignored, unless one is interested in
bound-bound transitions~\cite{Schunck08}. Then, the correlation
function decomposes into~\cite{Massignan}
\begin{equation}
 \Gamma_{\sigma\rightarrow e}(\omega_\mathrm{RF}) \propto
 \int\frac{d \bf k}{(2\pi)^4} A_\sigma(k,\xi_{k,\sigma}-\omega_\mathrm{RF})
 f(\xi_{k,\sigma}-\omega_\mathrm{RF}),
 \label{R}
\end{equation}
with $f(\omega)=[\exp(\beta\omega)+1]^{-1}$. The theory used here is described in detail in Refs.~\cite{Massignan,BruunBaym}.

It is useful to introduce a measure of the fraction of $|\sigma\rangle$ atoms bound in molecular states, which may have a finite or an infinite lifetime. These atoms may alternatively be thought of as bound in nonzero-momentum Cooper-pair fluctuations.
 For this quantity we introduce the definition $\Delta n_\sigma \equiv -\partial\Delta\Omega/\partial{\mu_\sigma}$, such that
$n_\sigma=n_\sigma^0+\Delta n_\sigma$ with $n_\sigma^0=\int d {\bf
k} f(\xi_{k,\sigma})/(2\pi)^{3}$.
 On the BEC side of the resonance, the many-body T matrix $T^{2B}/[1-T^{2B}\Xi(K,\epsilon+i0)]$ has a real pole at energy $\epsilon_K$.
 The contribution from these real poles to $\Delta n_\sigma$ can easily be shown to be
\begin{equation}
\Delta n_{\sigma}^{(\mathrm{s})}=\int \frac {d \bf
K}{(2\pi)^3}\frac{Z_{K,\sigma}}{\mathrm{e}^{\beta \epsilon_K}-1},
\label{Undamped}
\end{equation}
where $Z_{K,\sigma}=-\partial\epsilon_K/\partial{\mu_\sigma}$ and the
integral is over the values of $\mathbf{K}$ for which there is a real
pole. Far on the BEC side, $\Delta n_\sigma$ therefore gives the
density of stable molecules with energy $\epsilon_K$ and with an occupation probability given by $Z_{K,\sigma}$. For $a\rightarrow 0^+$,
$\epsilon_K=K^2/4m-1/ma^2-(\mu_\downarrow+\mu_\uparrow)$ and
$Z_{K,\sigma}=1$, such that we recover the result of $n_\sigma$
noninteracting molecules. However,
close to the unitarity limit there are no stable molecules. In
this regime, the distinction between bound and unbound atoms is
highly dependent on the experimental probe used to explore this
question and the many-body state simply cannot be thought of as
molecules mixed with atoms.

\begin{figure}
\includegraphics[angle=0,width=\columnwidth]{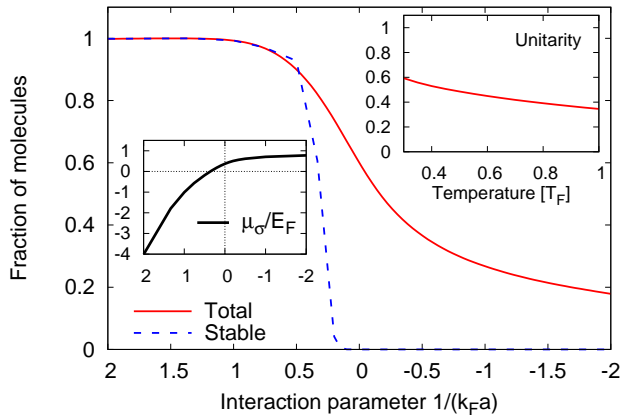}
\caption{\label{fig:molFractionBalanced} Fluctuation contribution $\Delta n_\sigma / n_\sigma$ and stable molecular fraction $\Delta n_{\sigma}^{(\mathrm{s})} / n_\sigma$ for a balanced gas at $T=0.3T_F$.
 Left inset: $\mu_\sigma/E_F$ vs. $1/(k_Fa)$ at $T=0.3T_F$. Right inset:  $\Delta n_\sigma / n_\sigma$ as a function of temperature at unitarity.}
\end{figure}
Consider first the unpolarized case ($n_\uparrow=n_\downarrow$) at $T=0.3T_F$. Here and in the
following $k_BT_F=E_F=k_F^2/2m=(6\pi^2n_\uparrow)^{2/3}/2m$ is
the Fermi energy of the majority component. In
Fig.~\ref{fig:molFractionBalanced}, we plot the fraction $\Delta
n_{\sigma}/n_{\sigma}$ of atoms which are bound into molecular states. In the BCS
regime, this number is small.
 The fraction increases continuously with increasing interaction toward the BEC regime.
 At unitarity, $\Delta n_\sigma \simeq 0.6n_\sigma$. On the BEC side of the resonance, the chemical potential becomes negative for $1/(k_Fa)\gtrsim0.3$, and correspondently the many-body T-matrix acquires a pole with vanishing imaginary part.
 This leads to a sharp increase in the number of atoms bound in \emph{stable} molecular states.
 In fact, for $1/(k_Fa)\simeq 0.5$ we find $\Delta n_{\sigma}^{(\mathrm{s})}>\Delta n_{\sigma}$, showing that, even if stable molecules exist, the physical picture of atoms either being free or bound in molecular states is too simple for the normal phase in the strongly-interacting limit~\cite{OhashiGriffin}.
 Of course, deep in the BEC regime all atoms are bound in molecular states and then
$\Delta n_\sigma=\Delta n_{\sigma}^{(\mathrm{s})}=n_\sigma$.
\begin{figure}
\includegraphics[angle=0,width=\columnwidth]{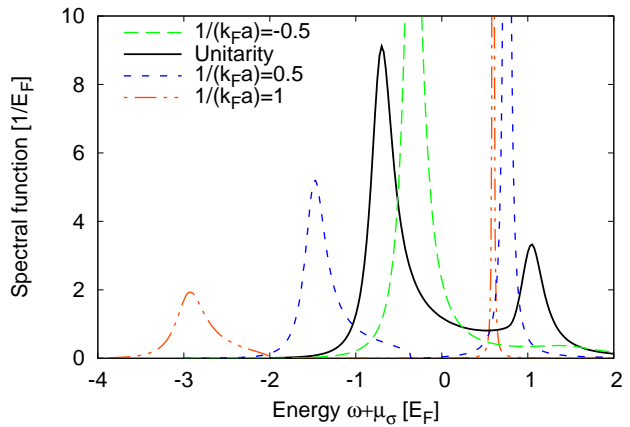}
\caption{\label{fig:spectralFunctionBalanced} Spectral functions
$A_\sigma(k=0,\omega)$ for a balanced gas at $T=0.3T_F$. The frequency axis is shifted by $\mu_\sigma$ such that the spectral function of an ideal gas
would be a delta function located at the origin.}
\end{figure}

\begin{figure}
\includegraphics[angle=0,width=\columnwidth]{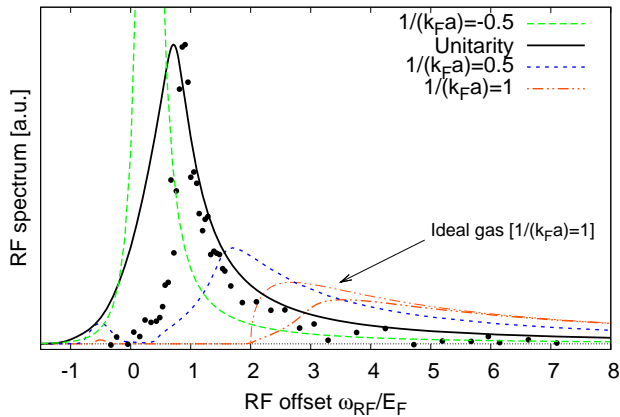}
\caption{\label{fig:spectrumBalanced} RF spectrum of a balanced
Fermi gas at $T=0.3T_F$. 
The molecular peak emerges for $1/(k_{F}a)\gtrsim0.3$. The dots are experimental points at unitarity and at $T=0.1T_F$ from \cite{Schunck08}.}
\end{figure}
In Fig.~\ref{fig:spectralFunctionBalanced}, we plot the spectral
functions for the balanced gas at $T=0.3T_F$ and in
Fig.~\ref{fig:spectrumBalanced} the corresponding RF spectrum for
various coupling strengths. The spectral functions with $k\lesssim k_F$, which give the main contribution to the RF signal in Eq.~(\ref{R}), look similar to the $k=0$ case plotted here. 
Deep in the BCS regime, the spectral function (RF spectrum)
displays a single narrow peak shifted to negative (positive)
energy, showing that the gas may be described as a Fermi mixture
with an attractive mean-field shift.
The spectral function in the unitarity limit has a double-peak structure characteristic of the pseudogap.
 As shown both experimentally and theoretically \cite{Shin,Massignan}, there is no corresponding double-peak structure in the RF spectrum, which has a single broad
peak shifted towards positive energy due to the effectively attractive interaction.
 In Fig.~\ref{fig:spectrumBalanced} we also show the experimental data
obtained in the unitarity limit~\cite{Schunck08}. We see a good
agreement between the theoretical and experimental results. The
discrepancy at low frequencies is most likely due to the theoretical curve being calculated at $T\simeq T_c$, while the 
experiment is performed at $T<T_c$ where the superfluid
nature of the gas results in a true gap in the RF
spectrum~\cite{Perali}.
 From Figs.~\ref{fig:molFractionBalanced} and \ref{fig:spectralFunctionBalanced} we see how the emergence of
stable molecules for $1/(k_Fa)\gtrsim 0.3$ leads to the spectral
function developing a gap between its two peaks.
 The peak at negative energy corresponds to the molecule continuum and can be understood as
follows. An atom in state $|\sigma\rangle$ with momentum
${\mathbf{k}}$ and energy $\omega$ can pair with an atom in state
$|-{\sigma}\rangle$ with momentum ${\mathbf{p}}$ and kinetic
energy $p^2/2m$ forming a molecule with energy
$E_b+K^2/4m$. Here $E_b$ and $\mathbf{K}={\mathbf{k}}+{\mathbf{p}}$ are respectively the binding
energy and the momentum of the molecule.  Energy conservation gives 
\begin{equation}
\omega=E_b+\xi_{k,\sigma}-\frac{({\mathbf{k}}-{\mathbf{p}})^2}{4m}.
\label{MolContin}
\end{equation}
In the BEC limit,  $A_\sigma(k,\omega)$
therefore consists of a molecular part for $w \le
E_b+\xi_{k,\sigma}$ separated by $E_b\simeq-1/ma^2$ from the atomic part. Importantly, this structure of the spectral
function translates into a two-peak structure in the RF
spectrum as shown explicitly in Fig.~\ref{fig:spectrumBalanced}.
The molecular part is now located at high frequencies as one needs
to provide additional energy to break the molecule.

In the BEC limit, the system can be considered as an ideal gas of
atoms mixed with tightly bound molecules with binding energy
$E_b=-1/ma^2$. The RF photons flip the $|\sigma\rangle$ atoms to
the final empty state $|e\rangle$ with a rate given by the Golden
Rule expression $\Gamma^\mathrm{a}_{\sigma\rightarrow
e}(\omega_\mathrm{RF}) \propto \int d\mathbf{k}
f(\xi_{k,\sigma})\delta(\omega_\mathrm{RF})/(2\pi)^3$. The rate coming from the
$|\sigma\rangle$ atoms bound in molecules is given by
\begin{multline}
\label{idealGasSpectrum} \Gamma^{m}_{\sigma\rightarrow e}(\omega_\mathrm{RF}) \propto
\int \frac{d\mathbf{k} d\mathbf{K}}{(2\pi)^6}|\varphi_\mathbf{k}|^2
n_\mathbf{K}
[1-f(\xi_{\mathbf{K}/2-\mathbf{k},-\sigma})]\\
\delta(\omega_\mathrm{RF}+K^2/4m+E_b
-\xi_{\mathbf{K}/2+\mathbf{k},e}-\xi_{\mathbf{K}/2-\mathbf{k},-\sigma})
\end{multline}
where $n_\mathbf{K}$ is the Bose distribution for the molecules.
Eq.~(\ref{idealGasSpectrum}) describes molecules with momentum ${\mathbf{K}}$ splitting into pairs of atoms in spin states $|e\rangle$ and $|-{\sigma}\rangle$ and with momenta $\mathbf{K}/2\pm\mathbf{k}$. The matrix
element for this process does not depend on ${\mathbf{K}}$ due to
Galilean invariance, and is equal to the molecule wavefunction, which is of the Yukawa form $\varphi_\mathbf{k}=\sqrt{8\pi a^3}/(1+k^2a^2)$.

The RF spectrum obtained from Eq.~(\ref{idealGasSpectrum}) with a
molecular binding energy $E_b/E_F=2$, corresponding to $1/(k_Fa)=1$,
is plotted with a thin line in Fig.~\ref{fig:spectrumBalanced}. 
There is good agreement between the ideal gas model and
the full many-body calculation for large $\omega$ since the vacuum
Yukawa wavefunction for the molecules is accurate for high
momenta. Closer to the threshold for molecular dissociation, the many-body RF spectrum is suppressed as compared to
the ideal gas prediction due to Pauli blocking effects; the  medium suppresses the components of the molecule wavefunction with momenta $k\lesssim k_F$. We conclude that these interesting medium effects on
the molecular wavefunction can be detected in RF spectroscopy.

\begin{figure}
\includegraphics[angle=0,width=\columnwidth]{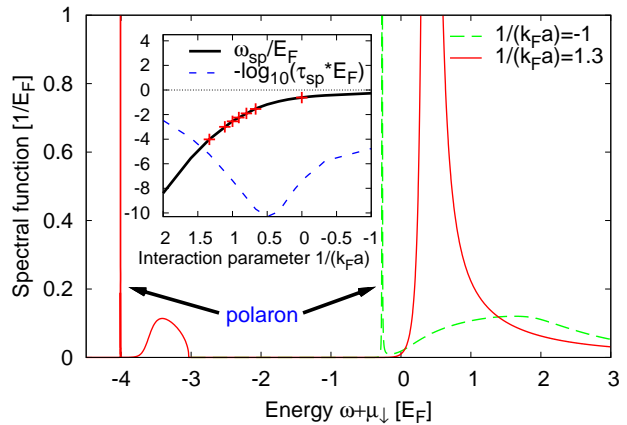}
\caption{\label{fig:spectralFunctionImbalanced}Spectral function
$A_\downarrow(k=0,\omega)$ for the minority component of an
extremely imbalanced gas at $T=0.05T_F$. Inset: energy $\omega_\mathrm{sp}$ and inverse lifetime $1/\tau_\mathrm{sp}$ of the spin-polaron. The red crosses are the
Monte-Carlo results from Ref.~\cite{ProkofevSvistunov}.}
\end{figure}
\begin{figure}
\includegraphics[angle=0,width=\columnwidth]{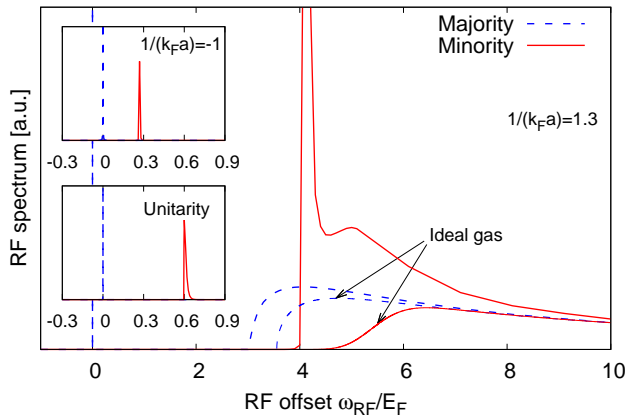}
\caption{\label{fig:spectrumImbalanced} RF spectrum of an extremely imbalanced gas at $T=0.05T_F$.
 The thin lines are the ideal gas results from Eq.~(\ref{idealGasSpectrum}).
 Most of the majority spectral weight resides in the delta-like peak at $\omega_\mathrm{RF}=0$. Only a fraction of order $1/N$ (where $N$ is the total number of particles) is
 in the high frequency shoulder.
}
\end{figure}

We proceed now to the analysis of a highly imbalanced gas
($n_\uparrow \gg n_\downarrow$) at $T=0.05T_F$. In
Fig.~\ref{fig:spectralFunctionImbalanced}, we plot the spectral
function of the minority component for $k=0$ (since $k_{F\downarrow}/k_F\ll 1$, where $k_{F\downarrow}=(6\pi^2n_\downarrow)^{1/3}$) on the BCS and BEC
sides of the Feshbach resonance. As for the balanced case, the spectral function for the minority component develops a broad
low-energy peak for $\omega<E_b$ due to the formation of molecules
as described in Eq.~(\ref{MolContin}). In addition, there is a
low-energy narrow peak which signals the presence of a
well-defined quasiparticle, the spin polaron. We find that this feature is present on both sides of the resonance, and its energy $\omega_\mathrm{sp}$ is in very good agreement with recent Monte-Carlo calculations~\cite{Lobo,ProkofevSvistunov}. Indeed, in the limit $n_\downarrow\ll n_\uparrow$ every diagram containing $|\downarrow\rangle$ holes is suppressed, and the ladder approximation accurately describes the main physical features of the system.
 The energy of the spin polaron is $-0.6E_F$ at unitarity, and on the BEC side it is  pushed down by the molecule continuum.
 Spin polarons arise due to the dressing of minority atoms by a large number of majority atoms. Indeed, no trace of those is present in the majority spectral function, which consists of a narrow peak at $\omega_\mathrm{RF}=0$ due to atoms and a molecule continuum for low energies.
 In the inset of Fig.~\ref{fig:spectralFunctionImbalanced} we also show that the lifetime $\tau_\mathrm{sp}\equiv-1/2\textrm{Im}\Sigma(k=0,\omega=\omega_\mathrm{sp})$
 of the spin polaron 
is very long close to unitarity. On the BCS side, the lifetime has a minimum (which at $T=0.05T_F$ lies at $k_Fa\simeq -1/3$), and in the limit $k_Fa\rightarrow 0^-$ it scales as 
$1/(k_Fa)^2$~\cite{BruunPethick}. The lifetime decreases moving toward the deep BEC side since the spin polaron  eventually  is absorbed in the molecule continuum and one ends up with an ideal gas of molecules and atoms. 

In Fig.~\ref{fig:spectrumImbalanced} we plot the RF
spectrum for the imbalanced case. To understand the rich physics, consider first the ideal gas model of atoms mixed with molecules introduced in Eq.~(\ref{idealGasSpectrum}) and given by the thin lines.
 The majority spectrum is characterized by a large noninteracting component, i.e., a narrow peak at zero offset, and a broad shoulder at $\omega_\mathrm{RF}\geq1/ma^2$, corresponding to molecule dissociation. The
minority spectrum displays no signal at zero offset, showing that
all $|\downarrow\rangle$ atoms are paired. Pauli blocking effects raise the threshold energy
for molecule break-up in the minority spectrum by roughly $E_F$.
 Indeed, when a RF photon breaks a molecule and transfers a $|\downarrow\rangle$ atom to the empty level $|e\rangle$, the remaining $|\uparrow\rangle$ atom cannot occupy states in the filled Fermi sea.
 At $T=0$ the photon therefore needs to carry at least $2E_F$ extra energy to transfer the $|\uparrow\rangle$ atom into a state above the Fermi sea, but nonzero temperatures reduce this threshold shift.

Consider now the results of the full many-body calculation shown
as thick lines in Fig.~\ref{fig:spectrumImbalanced}. In the $k_Fa\rightarrow0^+$ limit, our many-body spectra fully agree with
the ideal gas model of Eq.~(\ref{idealGasSpectrum}). Closer to the resonance on the BEC side, the many-body calculation agrees with the ideal atom-molecule mixture prediction for $\omega_\mathrm{RF}\gg |E_b|$, since the large-energy
$\omega_\mathrm{RF}^{-3/2}$ tails are due to two-body physics. However, interesting 
many-body effects appear near the molecular threshold. First, we
see from the majority spectrum that the medium increases the molecular energy above the two-body value $-1/ma^2$. As predicted by the ideal gas model, the minority threshold lies above the majority one. Furthermore, the qualitative new effect of the interactions, i.e., the presence
of the spin polaron, gives rise to a clear peak in the RF
spectrum. This leads to the important conclusion that one can
measure the energy of this intriguing state directly
by RF spectroscopy. 

We acknowledge useful discussions with A.\ Perali and C.\ H.\ Schunck, and thank B. Svistunov 
for kindly providing us with their Monte-Carlo data.


\begin{thebibliography}{99}
\bibitem{Regal}C.\ A.\ Regal \textit{et al.}, Nature \textbf{424}, 47 (2003).
\bibitem{Chin}C.\ Chin, Science \textbf{305}, 1128 (2004).
\bibitem{Schunck08}C. H. Schunck \textit{et al.}, arXiv:0802.0341v2.
\bibitem{Massignan}P.\ Massignan, G.\ M.\ Bruun, and H.\ T.\ C.\ Stoof,  Phys.\ Rev.\ A \textbf{77}, 031601(R) (2008).
\bibitem{Perali}A.\ Perali, P.\ Pieri, and G.\ C.\ Strinati, Phys.\ Rev.\ Lett.\ \textbf{100}, 010402 (2008).
\bibitem{Lobo}C.\ Lobo \textit{et al.}, Phys.\ Rev.\ Lett.\ \textbf{97}, 200403 (2006); R.\ Combescot  \textit{et al.}, \textit{ibid.} \textbf{98}, 180402 (2007).
\bibitem{Giorgini} S.\ Giorgini, L.\ P.\ Pitaevskii, and S.\ Stringari, arXiv:0706.3360.
\bibitem{NSR} P.\ Nozi\`{e}res and S.\ Schmitt-Rink,
J.\ Low.\ Temp.\  Phys.\ \textbf{59}, 195 (1985).
\bibitem{Pieri} P.\ Pieri, L.\ Pisani, and G.\ C.\ Strinati, Phys.\ Rev.\ B \textbf{72}, 012506 (2005).
\bibitem{Huliu}H. Hu, P.\ D.\ Drummond, and  X.-J. Liu, Nature Phys. \textbf{3}, 469 (2007).
\bibitem{Mueller}S. Basu and E. J. Mueller, arXiv:0712.1007.
\bibitem{BruunBaym} G.\ M.\ Bruun and G.\ Baym, Phys.\ Rev.\ A \textbf{74}, 033623 (2006).
\bibitem{OhashiGriffin} Y. Ohashi and A. Griffin, Phys. Rev. Lett. \textbf{89}, 130402 (2002).
\bibitem{Shin} Y. Shin \textit{et al.}, Phys. Rev. Lett. \textbf{99}, 090403 (2007).
\bibitem{ProkofevSvistunov} N. Prokof'ev and B. Svistunov, Phys.\ Rev.\ B \textbf{77}, 020408(R) (2008).
\bibitem{BruunPethick}G.\ M.\ Bruun and C.\ J.\ Pethick, Phys.\ Rev.\ Lett.\  \textbf{92}, 140404 (2004).
\end{thebibliography}
\end{document}